\documentclass[conference]{IEEEtran}
\usepackage[numbers]{natbib}
\usepackage{amsmath,amssymb,amsfonts}
\usepackage{algorithmic}
\usepackage{graphicx}
\usepackage{textcomp}
\usepackage{xcolor}
\usepackage{float}
\usepackage{booktabs}
\usepackage{bm}
\usepackage{url}
\usepackage{graphicx}
\def\BibTeX{{\rm B\kern-.05em{\sc i\kern-.025em b}\kern-.08em
    T\kern-.1667em\lower.7ex\hbox{E}\kern-.125emX}}
\begin{document}

\title{Diffusion models applied to skin and oral cancer classification\\
}

\author{\IEEEauthorblockN{José J. M. Uliana}
\IEEEauthorblockA{\textit{Labcin - Nature-inspired computing Lab} \\
\textit{Federal University of Espírito Santo}\\
Vitória, Brazil \\
ulianamjjorge@gmail.com}
\and
\IEEEauthorblockN{Renato A. Krohling}
\IEEEauthorblockA{\textit{Labcin - Nature-inspired computing Lab} \\
\textit{Graduate Program in Computer Science, PPGI, UFES} \\
\textit{Federal University of Espírito Santo}\\
Vitória, Brazil \\
krohling.renato@gmail.com}
}

\maketitle

\begin{abstract}
This study investigates the application of diffusion models in medical image classification (DiffMIC), focusing on skin and oral lesions. Utilizing the datasets PAD-UFES-20 for skin cancer and P-NDB-UFES for oral cancer, the diffusion model demonstrated competitive performance compared to state-of-the-art deep learning models like Convolutional Neural Networks (CNNs) and Transformers. Specifically, for the PAD-UFES-20 dataset, the model achieved a balanced accuracy of 0.6457 for six-class classification and 0.8357 for binary classification (cancer vs. non-cancer). For the P-NDB-UFES dataset, it attained a balanced accuracy of 0.9050. These results suggest that diffusion models are viable models for classifying medical images of skin and oral lesions. In addition, we investigate the robustness of the model trained on PAD-UFES-20 for skin cancer but tested  on the clinical images of the HIBA dataset. 
\end{abstract}


\section{Introduction}
\label{sec-intro}

Skin cancer, according to studies from the Global Cancer Observatory (GCO), had approximately 1,198,000 new cases worldwide in 2020, with non-melanoma skin cancer being the fifth most common cancer in terms of new cases, accounting for this high number \cite{gco2020}. In the same period, skin melanoma presented around 324,000 new cases. Together, these two types of cancer caused the death of about 120,000 people worldwide in 2020 \cite{gco2020}. Another type of cancer addressed in this work is mouth cancer. According to the GCO, in 2020, there were about 389,000 cases of Lip, Oral Cavity cancer, resulting in more than 188,000 deaths \cite{gco2020}.

Despite advances in skin and mouth cancer treatments, early diagnosis remains very important to reduce mortality. Cancer can manifest in various forms, making accurate diagnosis a challenge, especially in its early stages. Traditional diagnostic methods heavily rely on clinical examination and biopsy, which can be time-consuming and invasive. This highlights the need for innovative solutions that can help with early diagnosis, allowing for timely treatment and potentially reducing the number of fatalities.

There is a need to develop solutions that facilitate early diagnosis of skin and oral cancer since treatment in the early stages of the disease reducing the number of deaths. This has led to Computer-Aided Diagnosis (CAD) solutions, which aid in the early diagnosis of various diseases using medical imaging, such as clinical, histopathological, dermoscopic and x-ray images, and clinical data.

Traditionally, convolutional neural networks \cite{cnn, alexnet} are used for the classification of medical images \cite{pacheco2021attention, Oyedeji2024}. However, Transformer-based models \cite{vaswani2023attention} have been utilized for the classification of medical images, and comparative studies between convolutional neural networks and Transformers have emerged to verify whether a more modern approach addresses the limitations of convolutional neural networks, such as the lack of contextual awareness \cite{de_Lima_2022, PNDBUfes, DELIMA2023}. More recent works have utilized diffusion models \cite{ho2020denoising} for the classification of medical images, such as hyperspectral images of brain tissue \cite{sigger2024}.

This work conducts a study using datasets of clinical and histopathological images with the Dual-Guidance Diffusion Network for Medical Image Classification (DiffMIC) \cite{yang2023diffmic}, in order to compare with established models in the literature for the diagnosis of skin cancer using PAD-UFES-20 dataset \cite{PadUfes} and diagnosis of oral cancer using the P-NDB-UFES dataset \cite{NDBUfes}.

The main contributions of this work are twofold:
\begin{itemize}
\item we apply the Diffusion model (DiffMIC)  to two case studies involving diagnostic of 1) skin cancer using clinical images of the PAD-UFES-20 dataset and 2) oral cancer using histopathological images of the NDB-UFES dataset, in order to show its feasibility as compared to CNN and Transformers.
\item we investigate the performance of the DiffMIC model trained with the PAD-UFES-20 dataset when we test it on a similar dataset (HIBA) consisting of clinical image of skin lesion  in order to assess the robustness of the model.
\end{itemize}

The remaining sections of the paper is structured as follows: Section 2 presents the most relevant related works; Section 3 explains the Diffusion model (DiffMic) used to medical image classification; Section 4 presents and discuss the obtained results and Section 5 ends up the paper with conclusions as well as directions for future work.

\section{Related Works}
\label{sec-related-works}

Research in medical image classification typically focuses on Convolutional Neural Networks (CNNs). CNNs exhibit good performance in computer vision, but it lacks the capacity to embed the context of the data. Recently, there has been significant potential observed in replacing convolutional models by Transformer-based models, due to their ability of taking into account the context of the data through attention mechanisms.

Transformers \cite{vaswani2023attention} emerged as a solution for natural language processing (NLP) problems and achieved success in classification tasks through Transformer-based image classification networks \cite{dosovitskiy2021image,heo2021rethinking,xu2021coscale}. \citeauthor{bai2021transformers} \cite{bai2021transformers} demonstrated that Transformers are no more robust than CNNs when the training strategy of the CNN is the same as that of the Transformer.

Studies focused on the use of Transformers in the diagnosis of skin and oral cancer emerged, including comparative studies \cite{de_Lima_2022} that experimentally concluded that Transfomer-based models achieve state-of-the-art results, and can improve performance in skin lesion classification tasks. \citeauthor{DELIMA2023} \cite{DELIMA2023} compared both multimodal fusion approaches combining histopathologycal images, demographical and clinical data with histopathologycal images only, and as indicated in the obtained results, Transformer-based models have an advantage over CNN and improve the results using fusion with clinical data. In the task of classifying the NDB-UFES \cite{NDBUfes} dataset with 3 classes, with no fusion method employed, Transformers like PiT \cite{heo2021rethinking} and ViT \cite{dosovitskiy2021image} have been applied.

Recently, diffusion models have gained considerable attention are diffusion models \cite{ho2020denoising}. Initially employed in generative tasks \cite{rombach2022latent} and also in the context of medical images \cite{Franzes23}, diffusion models outperformed three well-established Generative Adversarial Network (GAN) based models on three datasets using metrics such as Fréchet Inception Distance (FID), Precision, and Recall.

\citeauthor{beatgan2023} \cite{beatgan2023} compared the performance in classification tasks of six different models, of which three are diffusion models. The diffusion model architecture used was an enhanced version of Guided Diffusion, and tests were conducted on the ImageNet-1K and ImageNet-50 datasets \cite{imagenet}. The results indicated an advantage of diffusion models over GAN models.

Diffusion models have also demonstrated favorable outcomes in data augmentation tasks. \citeauthor{trabucco2023effective} \cite{trabucco2023effective} proposed a data augmentation strategy based on a diffusion model called DA-Fusion and compared it with state-of-the-art strategies such as RandAugment \cite{cubuk2020randaugment} and Real Guidance \cite{he2022synthetic}. The evaluation comprehended seven datasets including Leafy Spurge dataset \cite{trabucco2023effective}. A consistent improvement in few-shot classification accuracy was observed with the use of DA-Fusion.

\citeauthor{azizi2023} \cite{azizi2023} compared several generative models, among GANs and diffusion models, with a custom model for image data augmentation trained on large multimodal dataset, the models have been used to augment ImageNet \cite{imagenet}. The augmented dataset was trained and tested using ResNet \cite{resnet}, ViT \cite{dosovitskiy2021image}, and DeiT-based classifiers, and resulted in improved classification accuracy across all model architectures.

\citeauthor{yang2023diffmic} \cite{yang2023diffmic} conducted a study on the application of diffusion models in the classification of medical images across three tasks. The PMG2000 dataset, an in-home dataset, was employed for placental maturity inference, the HAM10000 dataset \cite{Tschandl2018} for skin lesions, and the APTOS2019 dataset \cite{aptos2019-blindness-detection} for diabetic retinopathy. The authors developed the DiffMIC model, which was compared with CNN and Transformer-based models in three classification tasks, each corresponding to a specific dataset. In all cases, DiffMIC exhibited higher classification accuracy.

Despite the classification capabilities of diffusion models in medical images, DiffMIC  was not applied yet to clinical images of skin cancer and histopathological images of oral cancer \cite{yang2023diffmic}. In other words, its generalization capacity for this specific set of tasks was not verified. This study aims to address these gaps by conducting comparative analyses among diffusion models, CNNs, and Transformers in the realm of skin and oral cancer classification utilizing the PAD-UFES-20 \cite{PadUfes} and P-NDB-UFES \cite{NDBUfes} datasets, respectively.

\section{Diffusion Models}
\label{sec-diff-models}

\subsection{Introduction}

Diffusion model \cite{ho2020denoising} is based on the stochastic process of diffusion, concept used in the Brownian motion. Diffusion models emerged as the state-of-the-art for generative models \cite{dhariwal2021diffusion}. Nevertheless, this kind of model has been gaining importance in classification tasks in the last few years \cite{han2022card}.

The central idea behind diffusion models is that, through stochastic processes, that is, a set of random variables that change over time according to probabilistic rules, noise can progressively transform into an image or any other type of information one wishes to obtain. Training involves the diffusion process, in which noise is applied to the data, and the generative process, where typically a U-Net \cite{ronneberger2015unet}, is used to determine the noise that was applied during the diffusion process, thus allowing the process to be reversed.

Some frameworks also include a guidance process in the diffusion model, where a neural network can receive features from an encoder or embeddings derived from the attention mechanism. This allows textual prompts to generate images \cite{rombach2022latent}.

\subsection{Dual-Guidance Diffusion Network for Medical Image Classification (DiffMIC)}
\label{sec-exp-micdiff}

\begin{figure}[!h]
    \centering
    \includegraphics[width=0.5\textwidth]{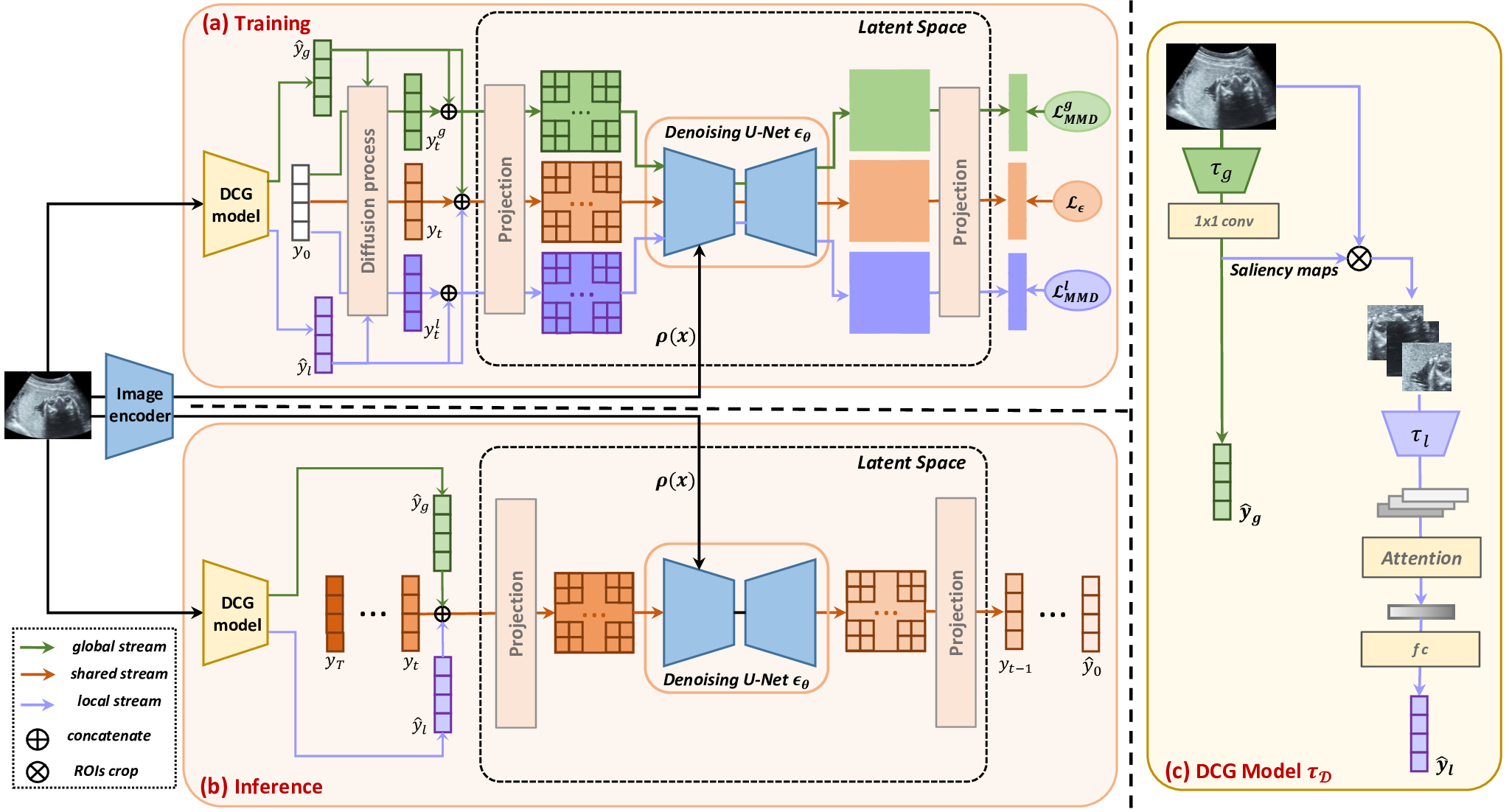}
    \caption{DiffMIC framework with training (a) and inference (b) pipelines. There is also a block containing the DCG model (c) pipeline. This figure was taken from the seminal work of DiffMIC \cite{yang2023diffmic}.}
    \label{fig:diffmic}
\end{figure}

\citeauthor{yang2023diffmic} \cite{yang2023diffmic} introduced diffusion models for medical image classification through DiffMIC using the concept of dual-granularity, which employs both local and global encoders to extract features through the Dual-granularity Conditional Guidance (DCG) strategy. The entire framework containing the training and inference pipelines, as well as the DCG model, is shown in Figure \ref{fig:diffmic}.

In the DCG model, represented as the rightmost block shown in Figure \ref{fig:diffmic}, the global prior is obtained from a global encoder, followed by a 1x1 convolution to create a saliency map. The local prior is generated from this map by identifying regions of interest (ROIs) to create image crops. Each crop is processed by a local encoder, and the outputs are combined into a weighted vector by an attention mechanism. The local prior is then obtained by passing this vector through a linear layer.

In the training phase, represented as the top block shown in Figure \ref{fig:diffmic}, the ground truth vector $y_0$ undergo the diffusion process considering the DCG priors. The results are concatenated with the DCG priors and projected into the latent space before undergoing noise removal, performed by a U-Net. The output from the generative process is used to calculate the loss for each encoder in DCG and for the noise estimator in U-Net.

In the inference process, represented as the bottom block shown in Figure \ref{fig:diffmic}, only the generative process is applied, with no diffusion step. The input to the U-Net is a vector created from the concatenation of the local and global priors, along with the output from the previous inference step, after projection into the latent space. The process is iterative, with the output of each stage becoming the input to the next, starting with Gaussian noise. The expected final output is a vector indicating the probabilities of the input belonging to a specific class.

DiffMIC takes into account the features $\rho(x)$ obtained through the image encoder to reconstruct the image using $\rho(x)$ as a guiding signal. Thus, $\epsilon_{\theta}$ receives $\rho(x)$ as a parameter. The shared stream loss function $L(\theta)$ used in DiffMIC is given by:
\begin{equation}
    L(\theta) := || \epsilon - \epsilon_{\theta}(\rho(x), y_t, \Hat{y}_g, \Hat{y}_l, t) ||^2
\end{equation}
where $y_t$ is the noisy one-hot encoded vector, $\Hat{y}_g$ is the global prior, and $\Hat{y}_l$ is the local prior. This equation is based on the Denoising Diffusion Probabilistic Models (DDPM) loss function \cite{ho2020denoising}, given by:
\begin{equation}
    L_{simple}(\theta) := \mathbb{E}_{t, x_0, \epsilon}\left[|| \epsilon - \epsilon_\theta(x_t, t) ||^2\right]
\end{equation}
where $\theta$ is the set of the model learned parameters, $t$ is the current time step, $x_0$ is the denoised latent and $x_t$ is the current noisy latent.

\section{Experimental Results}
\label{sec-exp}

We present the datasets and metrics used to evaluate the models, as well as the experimental methodology and settings. Next, the experimental results obtained from training DiffMIC using the PAD-UFES-20 and P-NDB-UFES datasets are presented and compared with results presented in the literature.

\subsection{Datasets}
\label{sec-exp-dataset}

The first dataset, PAD-UFES-20 \cite{PadUfes}, is used to assess the model's accuracy in the task of classifying clinical images of skin lesions. The second dataset, P-NDB-UFES \cite{NDBUfes}, is used to assess the accuracy of the model in the task of classifying histopathological images of oral lesions.

\subsubsection{PAD-UFES-20}
\label{sec:pad}

\begin{figure}[H]
    \centering
    \includegraphics[width=.5\textwidth]{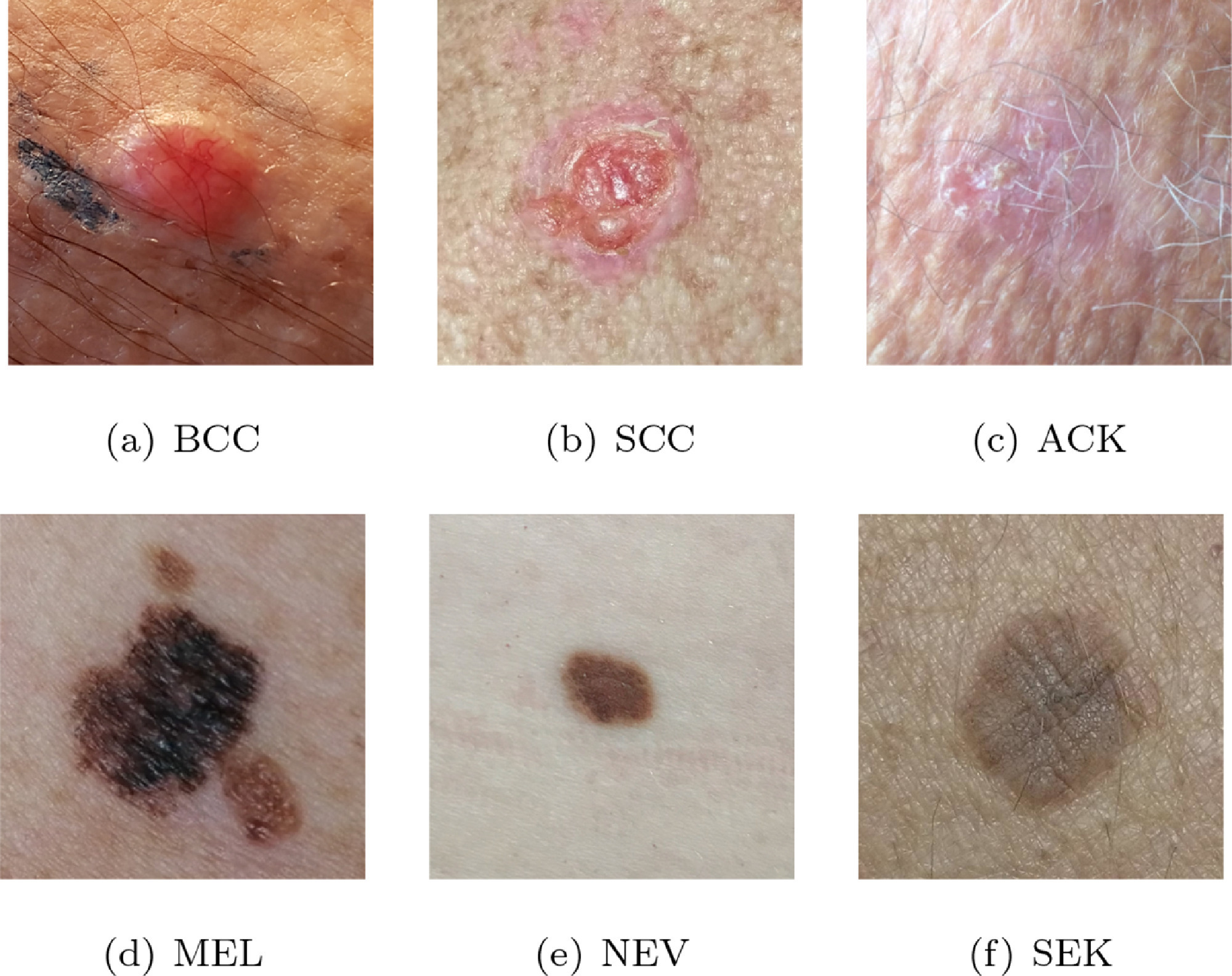}
    \caption{A sample of each type of skin lesion present in PAD-UFES-20 dataset. (a) Basal Cell Carcinoma of skin. (b) Squamous Cell Carcinoma. (c) Actinic Keratosis. (d) Malignant Melanoma. (e) Melanocytic Nevus of Skin. (f) Seborrheic Keratosis \cite{PadUfes}.}
    \label{fig:padufes20}
\end{figure}

The PAD-UFES-20 dataset \cite{PadUfes} comprises a total of 2,298 clinical images of skin lesions divided into 6 classes, where 3 are cancerous: melanoma (MEL), basal cell carcinoma (BCC) and squamous cell carcinoma (SCC), and 3 are benign: actinic keratosis (ACK), seborrheic keratosis (SEK) and nevus (NEV). The classes distribution is presented in Table \ref{tab:pad-train-sum}. Figure \ref{fig:padufes20} shows a sample image of each class.

\begin{table}[htbp]
\caption{Training and Testing Sets of PAD-UFES-20}
\begin{center}
\begin{tabular}{|c|c|c|c|c|}
\hline
\textbf{Type of} & \multicolumn{2}{|c|}{\textbf{ Train Set Samples}} & \multicolumn{2}{|c|}{\textbf{ Test Set Samples}} \\
\cline{2-5}
\textbf{Lesion} & \textbf{Count} & \textbf{Percentage} & \textbf{Count} & \textbf{Percentage} \\
\hline
ACK & 608 & 31.70 \% & 122 & 31.85\% \\
BCC & 704 & 36.62 \% & 141 & 36.81\% \\
MEL & 44 & 2.29 \% & 9 & 2.35\% \\
NEV & 204 & 10.73 \% & 40 & 10.44\% \\
SCC & 160 & 8.33 \% & 32 & 8.36\% \\
SEK & 196 & 10.19 \% & 39 & 10.18\% \\
\hline
\end{tabular}
\label{tab:pad-train-sum}
\label{tab:pad-test-sum}
\end{center}
\end{table}

For the purpose of evaluating the performance of the dataset, it was redivided into two classes: cancer and non-cancer. The first group includes images belonging to the classes MEL, BCC and SCC, while the second group includes images from the classes ACK, SEK and NEV.

\subsubsection{P-NDB-UFES}
\label{sec:pndb}

\begin{figure}[H]
    \centering
    \includegraphics[width=.5\textwidth]{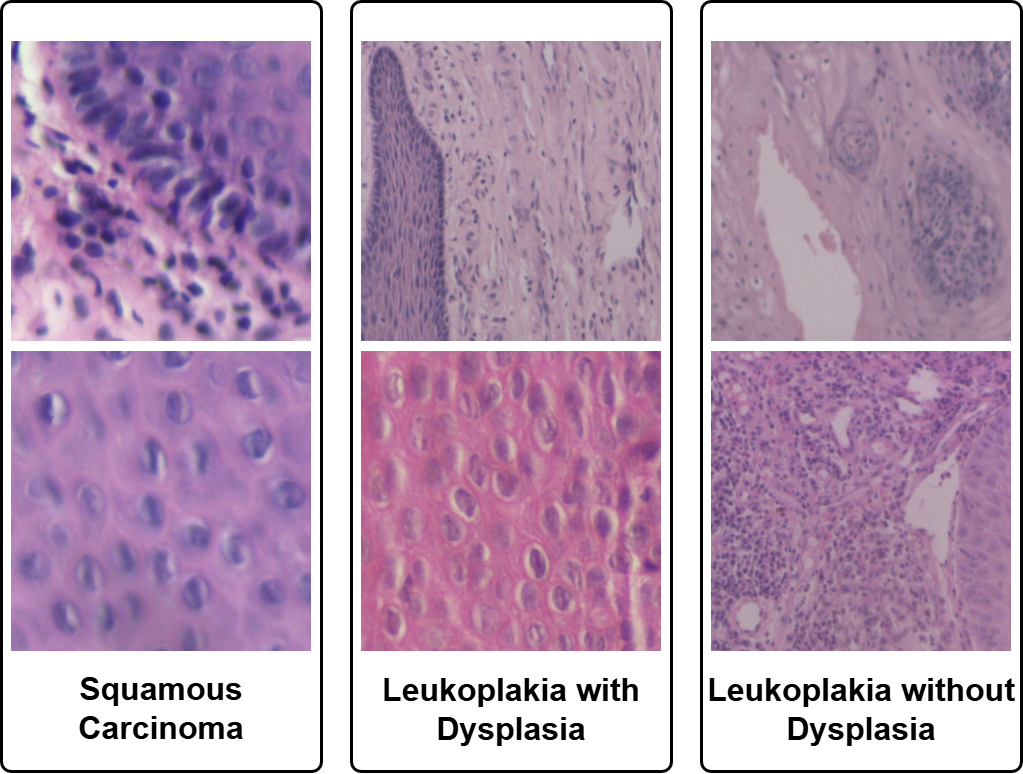}
    \caption{Samples of histopathological images from each class that exists in the P-NDB-UFES dataset \cite{DELIMA2023}.}
    \label{fig:NDBUfes}
\end{figure}

The NDB-UFES dataset \cite{NDBUfes} comprises a total of 237 histopathological images of squamous cell carcinoma (OSCC) and leukoplakia. The dataset is divided into three classes: leukoplakia with dysplasia, leukoplakia without dysplasia, and OSCC. From the 237 original images, 3,763 patches were extracted distributed by class as presented in Table \ref{tab:pndb-train-sum}. Since we used the patches and not the original images from the NDB-UFES dataset, as it is impractical to train a model with a dataset containing only 237 images, we will refer to the used dataset as P-NDB-UFES. Figure \ref{fig:NDBUfes} shows a sample image from each class.

\begin{table}[htbp]
\caption{Training and Testing Sets of P-NDB-UFES}
\begin{center}
\begin{tabular}{|c|c|c|c|c|}
\hline
\textbf{Type of} & \multicolumn{2}{|c|}{\textbf{Train Set Samples}} & \multicolumn{2}{|c|}{\textbf{Test Set Samples}} \\
\cline{2-5}
\textbf{Lesion} & \textbf{Count} & \textbf{Percentage} & \textbf{Count} & \textbf{Percentage} \\
\hline
OSCC & 939 & 29.94\% & 187 & 29.82\% \\
With dysplasia & 1608 & 51.28\% & 322 & 51.36\% \\
Without dysplasia & 589 & 18.78\% & 118 & 18.82\% \\
\hline
\end{tabular}
\label{tab:pndb-train-sum}
\label{tab:pndb-test-sum}
\end{center}
\end{table}

\subsection{Cross-validation}
\label{sec-projeto-met-kfold}
In order to obtain a fair classification metric comparison with other experiments \cite{pachecho2020,PNDBUfes, DELIMA2023}, a cross-validation strategy using K-Fold was employed. 

The datasets were split, where 1/6 was used for test and the other 5/6 were used for training. Each fold comprised 1/6 of the dataset. The folds for PAD-UFES-20 were generated by the script used in \citeauthor{de_Lima_2022} \cite{de_Lima_2022}, and for P-NDB-UFES, the folds provided in the Comma Separated Values (CSV) files accompanying the dataset were used. As a result of using this technique, the performance metrics for Task I are presented with a central value and an uncertainty (mean and standard deviation).

\subsection{Performance Metrics}
\label{sec-exp-metrics}

    The results were  analyzed considering the mean and standard deviation of the metrics in each combination. Four metrics widely used in the literature were used: Balanced Accuracy (BACC), Precision, Recall, and AUC are described by
    
    \begin{equation}
        BACC = \frac{\frac{TP}{TP + FN} + \frac{TN}{TN + TP}}{2}
        \label{BACC}
    \end{equation}
    
    \begin{equation}
        Precision = \frac{TP}{TP + FP}
        \label{PR}
    \end{equation}
    
    \begin{equation}
        Recall = \frac{TP}{TP + FN}
        \label{RE}
    \end{equation}
    
    \noindent where, the variables \textit{TP, TN, FP} and \textit{FN} represent True Positive, True Negative, False Positive, and False Negative, respectively.
    The AUC (Area Under the Curve) metric represents the area under the ROC (Receiver Operating Characteristic) curve. It measures the classifier's ability to distinguish between classes across all classification thresholds and is typically calculated using software tools rather than a simple formula involving TP, TN, FP, and FN. These metrics are used to evaluate the effectiveness of the algorithm used in the classification process.

\subsection{Experiment settings}
\label{sec-projeto-conf}

The hardware used during every task was a 13th generation Intel Core i9-13900K CPU, 128GB of RAM, three NVIDIA GeForce RTX 4090 graphic cards with a total of 72GB of VRAM. The code used to run the experiments is strongly based in the original DiffMIC repository \cite{yang2023diffmic}\footnote{https://github.com/scott-yjyang/DiffMIC} with some modifications to allow different datasets to be used, to employ the K-Fold cross-validation technique. In these settings, the process of training per epoch is performed in about 40 seconds for a total of 65 batches, in the case of the PAD-UFES-20 dataset.

The diffusion model used for training was maintained, in all experiments, the parameters defined in \citeauthor{yang2023diffmic} \cite{yang2023diffmic} for training on the HAM10000 dataset. With an input image size of 224x224 with 3 channels, the number of features in the U-Net is 6,144. The number of diffusion time steps $T = 1,000$, the forward process variances increase linearly from $\beta_1 = 10^{-4}$ to $\beta_T = 0.02$. The backbones for the DCG model is ResNet18, as well as the image encoder, this is done in order to keep the DiffMIC settings as close to the settings used in \citeauthor{yang2023diffmic} \cite{yang2023diffmic} as possible. The used criterion for the class prediction loss is the weighted cross entropy.

The optimizer Adam was utilized with a learning rate of 0.001, in contrast to \citeauthor{pacheco2021attention} \cite{pacheco2021attention} and \citeauthor{PNDBUfes} \cite{PNDBUfes} that used the SGD optimizer. This decision was made in order to obtain a competitive result in the DiffMIC implementation, once the results obtained using SGD were not satisfactory enough for the DiffMIC model (the accuracy did not surpass 50\% in the defined number of epochs). In accordance with 
\citeauthor{pacheco2021attention} \cite{pacheco2021attention} and \citeauthor{PNDBUfes} \cite{PNDBUfes} the model was trained for 150 epochs with a batch size of 30. Differently from what is done in \citeauthor{pacheco2021attention} \cite{pacheco2021attention} and \citeauthor{PNDBUfes} \cite{PNDBUfes} no early stop strategy is employed in the DiffMIC models training process, this is because, in the training sessions carried out using such a strategy, the accuracy obtained was considerably lower than that achieved without the use of early stopping.

\subsection{Results and Discussion}
\label{sub-sec-projeto-resultado-experimento1}

In the experiment, each of the datasets specified in section~\ref{sec-exp-dataset} was used to evaluate the DiffMIC model \cite{yang2023diffmic}. The Tables~\ref{tab-exp-i-pad} and~\ref{tab-exp-i-ndb} present the comparative results obtained from our experiments using DiffMIC and from other studies used as baseline for each experiment.

\begin{table}[htbp]
\caption{Results of the test using the PAD-UFES-20 dataset. All the non-diffusion models metrics were obtained from \citeauthor{pacheco2021attention} \cite{pacheco2021attention}.}
\begin{center}
\resizebox{\linewidth}{!}{%
\begin{tabular}{|c|c|c|c|c|c|}
\hline
\textbf{Model}         & \textbf{BACC}               & \textbf{Precision}          & \textbf{Recall}             & \textbf{F1}                 & \textbf{AUC}                \\ \hline
\multicolumn{6}{|c|}{\textbf{Diffusion Model}} \\ \hline
\textbf{DiffMIC}       & $0.6457 \pm 0.0430$         & $0.7206 \pm 0.0297$         & $0.7164 \pm 0.0346$         & $0.7153 \pm 0.0308$         & $0.8432 \pm 0.0250$ \\ \hline
\multicolumn{6}{|c|}{\textbf{Convolutional Neural Networks}} \\ \hline
EfficientNet-B4        & $0.6400 \pm 0.0290$         & -                           & -                           & -                           & $\bm{0.9110 \pm 0.0060}$         \\ 
DenseNet-121           & $0.6400 \pm 0.0420$         & -                           & -                           & -                           & $0.8930 \pm 0.0200$         \\ 
MobileNet-V2           & $0.6370 \pm 0.0180$         & -                           & -                           & -                           & $0.8980 \pm 0.0100$         \\ 
ResNet-50              & $0.6510 \pm 0.0500$         & -                           & -                           & -                           & $0.9010 \pm 0.0070$         \\ 
VGGNet-13              & $\bm{0.6540 \pm 0.0220}$    & -                           & -                           & -                           & $0.9010 \pm 0.0030$         \\ \hline
\end{tabular}}
\label{tab-exp-i-pad}
\end{center}
\end{table}

\begin{table}[htbp]
\caption{Results of the test using the P-NDB-UFES dataset. All the non-diffusion models metrics were obtained from \citeauthor{PNDBUfes} \cite{PNDBUfes}.}
\begin{center}
\resizebox{\linewidth}{!}{%
\begin{tabular}{|c|c|c|c|c|c|}
\hline
\textbf{Model}          & \textbf{BACC}                 & \textbf{Precision}           & \textbf{Recall}              & \textbf{F1-Score}            & \textbf{AUC}                \\ \hline
\multicolumn{6}{|c|}{\textbf{Diffusion Model}} \\ \hline
\textbf{DiffMIC}        & $0.9050 \pm 0.0066$ & $0.9144 \pm 0.0050$ & $0.9145 \pm 0.0050$ & $0.9144 \pm 0.0051$ & \textbf{$0.9432 \pm 0.0056$} \\ \hline
\multicolumn{6}{|c|}{\textbf{Convolutional Neural Networks}} \\ \hline
resnet50                & $0.9138 \pm 0.0118$         & $0.9177 \pm 0.0072$          & $0.9177 \pm 0.0072$          & $0.9177 \pm 0.0072$          & -                           \\ 
mobilenet               & $0.9149 \pm 0.0077$         & $0.9157 \pm 0.0061$          & $0.9157 \pm 0.0061$          & $0.9157 \pm 0.0061$          & -                           \\ 
densenet121             & \bm{$0.9191 \pm 0.0084$}    & \bm{$0.9193 \pm 0.0075$} & \bm{$0.9193 \pm 0.0075$} & \bm{$0.9193 \pm 0.0075$} & -                           \\ 
vgg16                   & $0.9109 \pm 0.0060$         & $0.9137 \pm 0.0064$          & $0.9137 \pm 0.0064$          & $0.9137 \pm 0.0064$          & -                           \\ \hline
\multicolumn{6}{|c|}{\textbf{Transformers}} \\ \hline
coat\_lite\_small       & $0.9084 \pm 0.0001$         & $0.9060 \pm 0.0001$          & $0.9060 \pm 0.0001$          & $0.9060 \pm 0.0001$          & -                           \\ 
pit\_s\_distilled\_224  & $0.8918 \pm 0.0001$         & $0.8920 \pm 0.0002$          & $0.8920 \pm 0.0002$          & $0.8920 \pm 0.0002$          & -                           \\ 
vit\_small\_patch16\_384 & $0.9042 \pm 0.0002$        & $0.9060 \pm 0.0002$          & $0.9060 \pm 0.0002$          & $0.9080 \pm 0.0001$          & -                           \\ \hline

\end{tabular}}
\label{tab-exp-i-ndb}
\end{center}
\end{table}

\subsubsection{Training and evaluating on PAD-UFES-20}

In order to test the performance of the DiffMIC model for clinical images of skin lesions, the model was trained using the PAD-UFES-20 dataset.

As presented in Table \ref{tab-exp-i-pad}, the DiffMIC model, trained on the PAD-UFES-20 training set and tested on its test set, achieved an average balanced accuracy of 64.57\%, placing it in a competitive position compared to CNN models, with the best model, VGGNet-13, achieving 65.40\% for the same metric.

\begin{figure}[H]
    \centering
    \includegraphics[width=.375\textwidth]{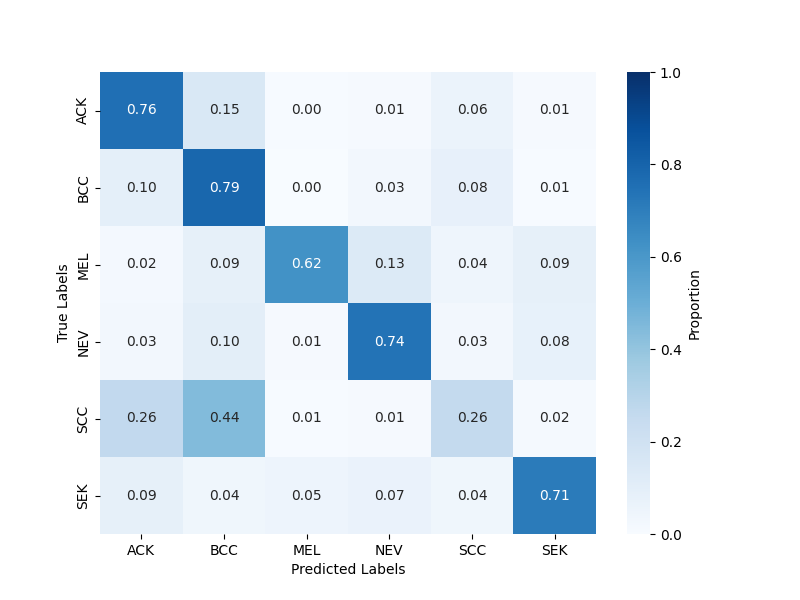}
    \caption{Confusion matrix obtained by evaluating DiffMIC on the PAD-UFES-20 test set.}
    \label{fig:cm-pad-test}
\end{figure}

The confusion matrix shown in Figure \ref{fig:cm-pad-test} was generated from all detections made by the 5 models trained on each of the 5 splits, using the K-Fold methodology. Each value in the matrix represents the proportion of samples from a specific class (each row corresponds to a ground truth class) that were classified as another class (each column corresponds to a predicted class). The matrix is row-normalized, meaning the sum of values in each row equals 1. The values in the confusion matrix in Figure \ref{fig:cm-pad-test} indicate that the model achieved good results in most classes, except for the SCC class, which was misclassified as the BCC class approximately 44\% of the time.

During the development of this work, the HIBA dataset \cite{RicciLara2023} was published and it also contains clinical and dermoscopic images publicly available captured at the Hospital Italiano de Buenos Aires between 2019 and 2022. HIBA consists of 346 clinical images, from a total of 1,270 images, including both clinical and dermoscopic, distributed across 10 classes. This makes it impractical to use for training DiffMIC due to the insufficient amount of clinical images. However, we use its clinical images to test DiffMIC's ability to generalize on clinical images of skin lesions.

It was necessary to remove 4 out of the 10 classes present in HIBA, keeping only the 6 that exist in both HIBA and PAD-UFES-20, ensuring the test remains consistent. Next, a test was conducted to verify the feasibility of using DiffMIC in separating the classes into cancer and non-cancer. This process was done for both PAD-UFES-20 and HIBA. Therefore, there are two pairs: one with 6 classes and another with 2 classes.

DiffMIC obtained an average balanced accuracy of 83.57\%, precision of 83.54\%, recall of 83.57\% and F1-Score of 83.52\% when trained and tested on the PAD-UFES-20, using 2,298 clinical images, divided into 2 classes. In comparison, \citeauthor{krohling2021smartphone} \cite{krohling2021smartphone} conducted the training and evaluation of the ResNet-50 CNN using the PAD-UFES-20 dataset, containing 2,057 clinical images, divided into 5 folds and distributed across 2 classes. The study explored different configurations to address class imbalance in the dataset. The model that achieved the highest average balanced accuracy reached 88.23\%, the same model achieved precision of 75.80\% and recall of 90.74\%.

The model trained on PAD-UFES-20 was then tested on the 309 images from HIBA, divided among the 6 classes that coexist in both HIBA and PAD-UFES-20. The results of this test is presented in Table \ref{tab-results-classes}. For the dataset with 2 classes, containing the same 309 images but with the classes mapped to cancer and non-cancer, the model was trained using a sub-set of PAD-UFES-20 also mapped to 2 classes \cite{PadUfes} and tested in the 309 images from HIBA. The result for the 2-class test is listed in the second row of Table \ref{tab-results-classes}.

\begin{table}[htbp]
\caption{Results of the test on the HIBA dataset with 6 and 2 classes. The model was trained on the PAD-UFES-20 dataset.}
\begin{center}
\resizebox{\linewidth}{!}{%
\begin{tabular}{|c|c|c|c|c|c|}
\hline
\textbf{N. of Classes} & \textbf{BACC}               & \textbf{Precision}          & \textbf{Recall}             & \textbf{F1-Score}           & \textbf{AUC}                \\ \hline
6                          & $0.4632 \pm 0.1241$        & $0.5407 \pm 0.1096$         & $0.4445 \pm 0.1645$         & $0.4423 \pm 0.1610$         & $0.7205 \pm 0.0825$         \\ 
2                          & $0.7068 \pm 0.0719$        & $0.7894 \pm 0.0330$         & $0.6236 \pm 0.1144$         & $0.6259 \pm 0.1144$         & $0.7268 \pm 0.0658$         \\ \hline

\end{tabular}}
\label{tab-results-classes}
\end{center}
\end{table}

\begin{figure}[H]
    \centering
    \includegraphics[width=.375\textwidth]{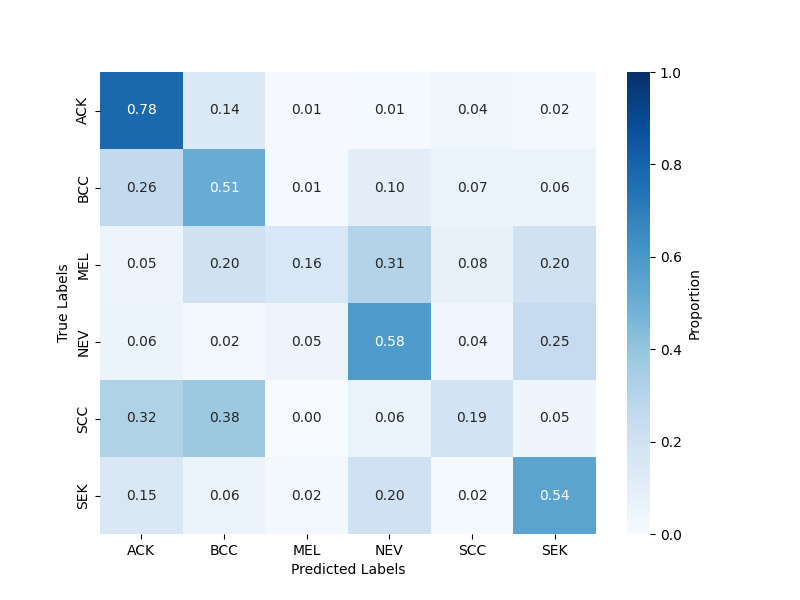}
    \caption{Confusion matrix obtained by evaluating DiffMIC trained using the 6 classes of the PAD-UFES-20 and tested on the clinical images of HIBA dataset.}
    \label{fig:cm-hiba-test}
\end{figure}

For the model trained with the 6 classes of PAD-UFES-20 and tested on the 309 images of HIBA mapped to the PAD-UFES-20 classes, the results were not satisfactory. Based on the average balanced accuracy of 46.32\% obtained, it indicates that the model is no better than random guessing. Figure \ref{fig:cm-hiba-test} shows the confusion matrix for the model trained with the 6 classes of PAD-UFES-20 and tested on the 309 images of HIBA mapped to the PAD-UFES-20 classes. When examining this confusion matrix, it becomes clear that the model only achieved good accuracy for the ACK class. However, when evaluating images belonging to the MEL and SCC classes, it made errors more than 80\% of the time.

\begin{figure}[H]
    \centering
    \includegraphics[width=.375\textwidth]{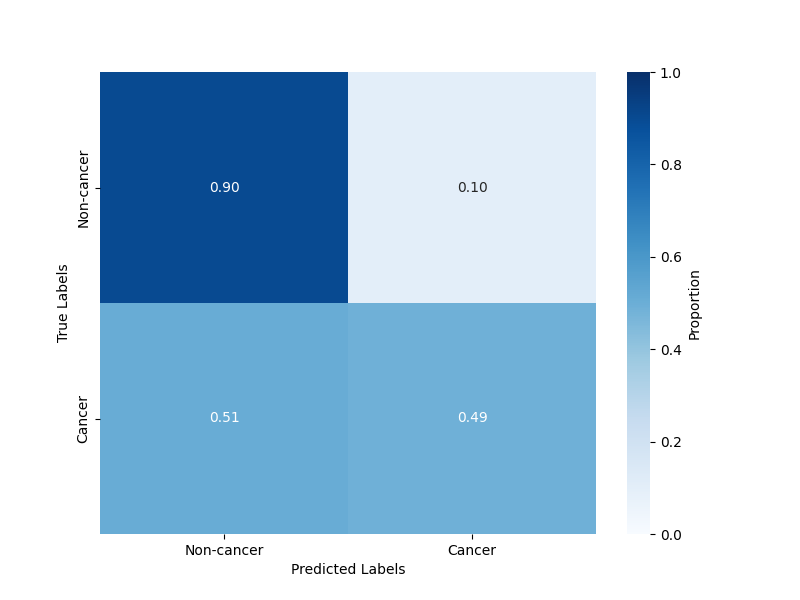}
    \caption{Confusion matrix obtained by evaluating DiffMIC trained using the cancer x non cancer classes of the PAD-UFES-20 and tested on the clinical images of the HIBA dataset.}
    \label{fig:cm-hibabin-test}
\end{figure}

For the model trained on PAD-UFES-20 with the 2 classes and tested on HIBA with the same 2 classes, cancer and non-cancer, the results were better, as presented in Table \ref{tab-results-classes}, with an average balanced accuracy of 70.68\%. Figure \ref{fig:cm-hibabin-test} shows the confusion matrix for the model trained with 2 classes in the PAD-UFES-20 and tested on the 309 images of HIBA mapped to the same 2 classes. The values in the confusion matrix indicate that the model achieved 90\% accuracy in predictions for the non-cancer class but made errors in 51\% of the predictions for the cancer class.

The Table \ref{tab-exp-i-pad} indicates that DiffMIC performed well on the PAD-UFES-20 dataset. However, when tested for its generalization capability, it did not generalize as well, particularly when using the 6 classes from PAD-UFES-20. Similar results were observed in analogous tests using CNNs \cite{Oyedeji2024}. There are many possible reasons for that, such as differences in ambiental conditions and different smartphone camera used during the image acquisition.

\citeauthor{Oyedeji2024} \cite{Oyedeji2024} conducted a similar experiment using CNN architectures, ResNet18, DenseNet169, MobileNetV3, and EfficientNetB7. For the 309 clinical images from the HIBA dataset as the test set, the highest balanced accuracy achieved was 43.6\%, using MobileNetV3, when trained on all six classes of the PAD-UFES-20 dataset. In a binary classification task, using 346 clinical images from the HIBA dataset as the test set and the PAD-UFES-20 dataset as the training set, ResNet18 achieved the best accuracy of 66.8\%.

The results found in \citeauthor{Oyedeji2024} \cite{Oyedeji2024}, along with those obtained experimentally in the present work, indicate that deep learning models, i.e., CNNs or diffusion models, within the domain of clinical image classification using the PAD-UFES-20 and HIBA datasets, are still not robust enough. 

\subsubsection{Training and evaluating on P-NDB-UFES}

In order to evaluate the performance of the DiffMIC model for histopathological images of oral lesions, the model was trained using the P-NDB-UFES dataset.

The obtained average balanced accuracy of 90.50\% indicate that for the P-NDB-UFES dataset, DiffMIC achieved competitive results, as presented in the Table \ref{tab-exp-i-ndb}, but it did not surpass the average balanced accuracy of any CNN model evaluated in \citeauthor{PNDBUfes} \cite{PNDBUfes}. The hybrid CNN-Transformer model CoaT achieved the average balanced accuracy of 90.84\% and the CNN DenseNet obtained the average balanced accuracy of 91.91\%. 

\begin{figure}[H]
    \centering
    \includegraphics[width=.375\textwidth]{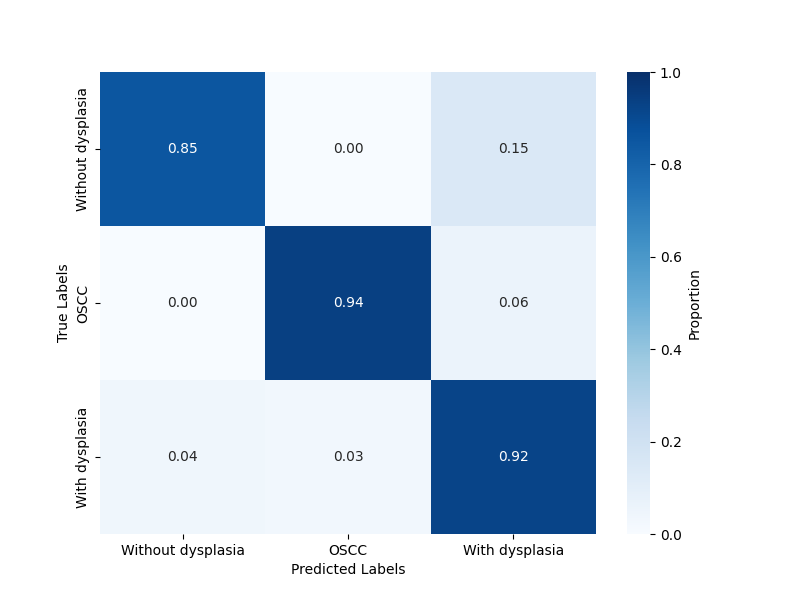}
    \caption{Confusion matrix obtained by evaluating DiffMIC using the test set of the P-NDB-UFES dataset.}
    \label{fig:cm-pndb-test}
\end{figure}

Figure \ref{fig:cm-pndb-test} shows the confusion matrix obtained from the test of the P-NDB-UFES dataset in the trained DiffMIC model. The values in the confusion matrix indicate that the model achieved good results across all classes, with the only notable exception being that it misclassified the~"without dysplasia"~class as~"with dysplasia"~approximately 15\% of the time.

\section{Conclusions}
\label{sec-conclusoes}

This study evaluates the DiffMIC model effectiveness in medical image classification, focusing on the PAD-UFES-20 and P-NDB-UFES datasets. On PAD-UFES-20, DiffMIC achieved a balanced accuracy of 64.57\%, comparable to established Convolutional Neural Networks (CNNs) models like EfficientNet-B4, DenseNet-121, MobileNet-V2, ResNet-50, and VGGNet-13. However, when tested on the HIBA dataset, the model's performance dropped to 46.32\% for six-class classification, indicating limited generalization. This improved to 70.68\% in binary classification for the case of cancer vs. non-cancer. For the P-NDB-UFES dataset, DiffMIC achieved a balanced accuracy of 90.50\%, closely matching DenseNet's 91.91\% and CoaT's 90.84\%, demonstrating its competitiveness with CNN and Transformer models. Future work may explore diffusion models for data augmentation to improve CNN and Transformer model performance.

\bibliographystyle{plainnat}
\bibliography{main}
\vspace{12pt}

\end{document}